\documentclass[aps,prb,twocolumn,showpacs,floatfix]{revtex4}

\usepackage{graphicx}
\graphicspath{{figs/}}
\begin{document}
\title{Mechanism of phonon localized edge modes}
\author{Jin-Wu~Jiang}
    \affiliation{Department of Physics and Centre for Computational Science and Engineering,
             National University of Singapore, Singapore 117542, Republic of Singapore }
\author{Jian-Sheng~Wang}
    \affiliation{Department of Physics and Centre for Computational Science and Engineering,
                 National University of Singapore, Singapore 117542, Republic of Singapore }

\date{\today}
\begin{abstract}
The phonon localized edge modes are systematically studied, and two conditions are proposed for the existence of the localized edge modes: (I) coupling between different directions ($x$, $y$ or $z$) in the interaction; (II) different boundary conditions in three directions. The generality of these two conditions is illustrated by different lattice structures: one-dimensional (1D) chain, 2D square lattice, 2D graphene, 3D simple cubic lattice, 3D diamond structure, etc; and with different potentials: valence force field model, Brenner potential, etc.
\end{abstract}

\pacs{63.20.Pw, 63.22.-m, 63.20.D-}
\maketitle

\pagebreak

In the lattice dynamics of a solid state system with $N$ atoms, there is one eigen vibrational vector $\vec{u}=(\vec{u}_{1}, \vec{u}_{2},..., \vec{u}_{N})$ corresponding to each normal phonon mode. In some phonon modes, a particular atom $i$ has large vibrational amplitude $u_{i}$, and the vibrational amplitude decays exponentially from this atom to its neighboring atom $j$,
\begin{eqnarray}
u_{j} \propto e^{-r_{ij}/L_{c}},
\label{eq_LEM}
\end{eqnarray}
where $r_{ij}$ is the distance between atoms $i$ and $j$, and $L_{c}$ is a constant. This is the so-called localized mode.\cite{Born} The constant $L_{c}$ can be taken as the localization length or penetration depth, and atom $i$ as the core of this localized mode. The localized mode usually appears around impurities with smaller mass\cite{Born}, vacancy defect\cite{Castro} or in some superlattice structures\cite{Dobrzynski, Sedrakyan, Venkatasubramanian}. Among others, there is a particular important class of localized mode, where the localization origins from the edge/surface configuration and the core is the the edge/surface atoms. The vibrational amplitude in this mode is very large for atoms on the edge/surface, and decays exponentially into the inner region of the system. This type of localized mode will be referred to as localized edge modes (LEM) in the present paper. The LEM exhibits its importance in various physical processes. It governs the generation of defects in the quench dynamics of a system driven across a quantum critical point.\cite{Bermudez} In thermal transport, LEM can localize the thermal energy on the edge/surface of the structure, which leads to very poor efficiency of pumping thermal energy into the system through these edge regions.\cite{Jiang} In the nonequilibrium Green's function scheme, LEM corresponds to delta peaks in the self energy function, which introduces much difficulty to implement it on computer by molecular dynamics.\cite{Wang} Very recently, it was found that the LEM also plays an important role in the dynamic instability of microtubules which are self-assembled hollow protein tubes playing important functions in live cells.\cite{Prodan} Besides the above listed importance, the LEM may have some possible applications in thermal nanodevices considering its ability to highly concentrate thermal energy. Although the importance of the LEM has been recognized in many fields, to the best of our knowledge, research on the origin of the LEM is still lacking, which is the aim of the present work.

Since LEM is a double-edged sword, it is important to understand its origin and find conditions for it, no matter we want to make use of its advantage or to avoid its fatal disadvantage. In this paper, we propose two conditions for the existence of the LEM: (I) coupling between different directions ($x$, $y$ or $z$) in the interaction; (II) different boundary conditions (BC) in three directions. The generality of these two conditions is displayed by different lattice structures: one-dimensional (1D) chain, 2D square lattice, 2D graphene, 3D simple cubic, and 3D diamond lattice. These two conditions are also valid where the interaction of the system is described by different potentials.

Two different interaction potentials are applied: the valence force field model (VFFM) and the Brenner interaction potential.\cite{Brenner}. The VFFM includes both longitudinal stretching and transverse bending types of energy\cite{SaitoR}:
\begin{eqnarray}
V_{l}&=&\frac{k_{l}}{2}[(\vec{u}_{j}-\vec{u}_{j})\cdot(\vec{e}_{ij}^{l})]^{2},\nonumber\\
V_{\perp}&=&\frac{k_{\perp}}{2}[(\vec{u}_{j}-\vec{u}_{j})\cdot(\vec{e}_{ij}^{\perp})]^{2},
\end{eqnarray}
where $k_{l}$ and $k_{\perp}$ are the corresponding force constants. The two unit vectors $\vec{e}_{ij}^{l}$ and $\vec{e}_{ij}^{\perp}$ are in the longitudinal and perpendicular directions between atoms $i$ and $j$. These two potentials are implemented in the ``General Utility Lattice Program".\cite{Gale} The force constant matrix is obtained from this code and diagonalized to achieve the eigen value, i.e the phonon frequency, and the eigen vector, i.e the vibrational amplitude vector $\vec{u}$. The localized or non-localized character of each mode is then determined by examining $\vec{u}$ with Eq.~(\ref{eq_LEM}).

As a fundamental property of localized mode, one particular atom vibrates with very large amplitude in this type of phonon mode. At the meantime, only those atoms in the neighborhood of this particular atom can feel its vibration and follow it to vibrate.\cite{Born} In case of no coupling between different directions in the interaction, there will be only the longitudinal vibration modes. These phonon modes prefer to travel along its vibrational direction instead of been localized.
\begin{figure}[htpb]
  \begin{center}
    \scalebox{1.1}[1.0]{\includegraphics[width=7cm]{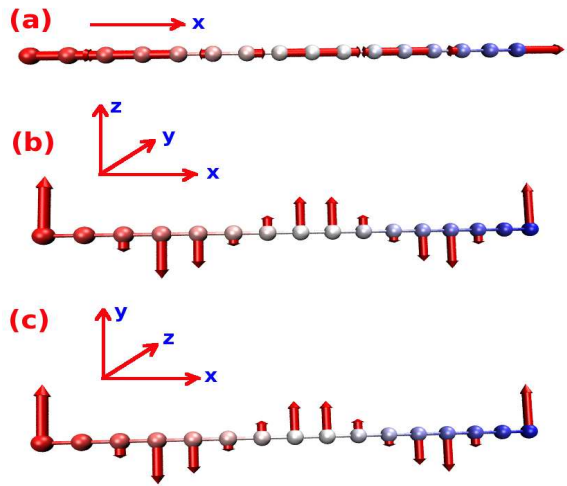}}
  \end{center}
  \caption{(Color online) Three typical extended phonon modes with large vibrational amplitude for the first boundary atom in a 1D chain.}
  \label{fig_chain}
\end{figure}
 So the first condition for LEM is (I): coupling between different directions in the interaction. The LEM is not simply a common localized mode, its localization core is exactly at the edge of the system. As a result, it will be sensitive to the local environment of the edge, which is determined by the BC. So the second condition for the LEM is (II): different BC in three directions. In the following, we are going to use different lattice structures to demonstrate these two conditions for the LEM. Before carrying out the calculation, we would like to make clear three issues. (1). For 2D or 3D structures, there are more than one atom on the boundary of the system. All of these boundary atoms are possible to serve as the localization core of the LEM. However, we find that it is adequate to treat just one of these boundary atoms, since all of them have equivalent number of LEM if there is any LEM in the system. For simplicity and clarity, we consider the LEM with the first atom on the left boundary as the localization core. (2). We stick to the Cartesian coordinate, where the $x$ axis is from left to right in the horizontal direction. The $y$ and $z$ axes will be rotated to display the structure more clearly where it is necessary. The $x$ and $y$ axes lie in the 2D plane in case of 2D lattice structure. (3). Each vibrational vector is demonstrated by an arrow (red online) plotted by VMD.\cite{HumphreyW} The direction of the arrow denotes the direction of the vibrational vector, while the length of the arrow is proportional to the value of the vibrational amplitude. In some figures for the LEM, there is a long arrow pointing from edge into center. Attention should be paid that this arrow displays the large vibrational amplitude of the boundary atom, not the vibration of the inner atoms.

\begin{figure}[htpb]
  \begin{center}
    \scalebox{1.1}[1.0]{\includegraphics[width=7cm]{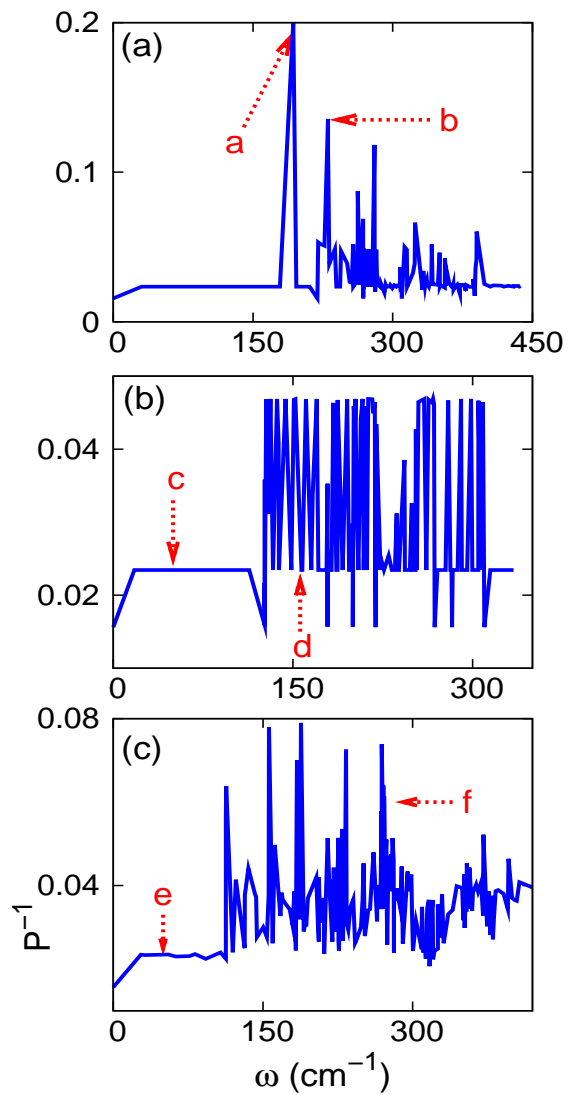}}
  \end{center}
  \caption{(Color online) The inverse participation ratio for 2D square lattice structure. (a) Two conditions are fulfilled; (b) condition (I) is broken; (c) condition (II) is broken. The labels a - f denotes the six phonon modes shown in Figure~\ref{fig_rectangle}.}
  \label{fig_rectangle_ipr}
\end{figure}
The inverse participation ratio (IPR) is a proper criteria for the phonon localization property.\cite{BellRJ} The IPR for phonon mode $\textbf{k}$ is defined through the normalized eigenvector $\textbf{u}_{k}$:
\begin{eqnarray}
P^{-1}_{k}=\sum_{i=1}^{N}\left(\sum_{\alpha=1}^{3} u_{i\alpha,k}^{2}\right)^{2},
\label{eq_ipr}
\end{eqnarray}
where $N$ is the total number of atoms. We can determine the value of IPR for the localized modes. It can be assumed that there are $m$ atoms that can vibrate in the localized mode, so the vibrational amplitude for each atom is $u=1/\sqrt{m}$ considering the normalization of eigen vector $\vec{u}$. Then the IPR for this mode can be obtained as $P^{-1}=1/m$. If $m=1$, we get a absolute localized mode with $P^{-1}=1$. This means that only one atom can vibrate in this mode.
\begin{figure}[htpb]
  \begin{center}
    \scalebox{1.3}[1.2]{\includegraphics[width=7cm]{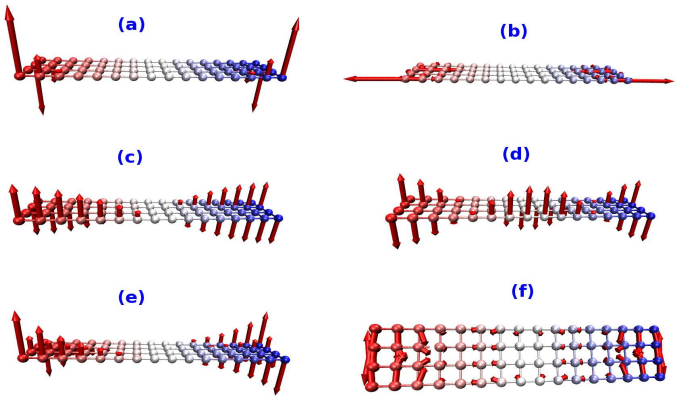}}
  \end{center}
  \caption{(Color online) (a) and (b) are two LEM in the 2D square lattice structure. (c) and (d) are two non localized modes because the condition (I) is broken. (e) and (f) are two non localized modes due to the broken of the condition (II).}
  \label{fig_rectangle}
\end{figure}
 We can say that this mode is absolute localized around this vibrating atom. We can also calculate the IPR for the translational phonon mode, where all atoms have the same vibrational amplitude $u=1/\sqrt{N}$. It is easy to see that the IPR for this mode is $P^{-1}=1/N$. As a result, it is straightforward to see the localization property of a phonon mode through its IPR value. If IPR for this mode is on the order of $1/m$, this mode is localized. Otherwise, if IPR is on the order of $1/N$, this mode is not localized. As we will see in the following, the value of $1/N$ is usually several orders smaller than $1/m$. So the IPR is an efficient criteria for the phonon localization.

Fig~\ref{fig_chain} is a 1D chain, where only one atom sits at the left/right boundaries. We find that there is no LEM in this system. In the figure we show three typical phonon modes in this system. In these modes, it is indeed the case that the boundary atom has large vibrational amplitude, however, the vibrational amplitudes do not decay from the edge into the center. So these kinds of phonon modes are none localized modes. We have applied open BC in the two boundaries. If we use the periodic BC, we will obtain the textbook harmonic oscillation eigen modes. Here we use the VFFM vibrational potential, where the interaction range is adjustable. If the interaction range is: $a<r_{cut}<2a$ ($a$ as the lattice constant), each atom only interacts with its first-nearest-neighboring (FNN) atoms; if $r_{cut}>2a$, the interaction is extended to the second-nearest-neighboring (SNN) atoms. We have changed the interaction range, and there is no LEM in all situations. This result is quite understandable. Because in the 1D system, there is no coupling between different directions in the interaction, and the edge only exists in the $x$ direction. As a result, neither conditions (I) nor (II) can be satisfied.

Fig~\ref{fig_rectangle_ipr} is the IPR for a 2D square lattice system. In panel (a),
\begin{figure}[htpb]
  \begin{center}
    \scalebox{1.3}[1.2]{\includegraphics[width=7cm]{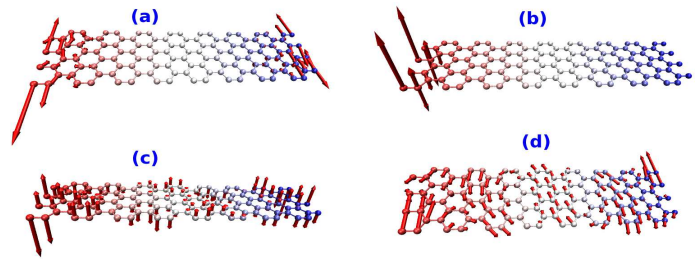}}
  \end{center}
  \caption{(Color online) LEM (a), (b), and non-localized modes (c), (d) in the graphene sheet. The non-localized mode is due to the broken of the condition (II). The vibrational potential is the VFFM.}
  \label{fig_graphene}
\end{figure}
 the two conditions for the LEM are satisfied in this system by considering interactions up to SNN which leads to the coupling between $x$ and $y$ directions; and applying the open/periodic BC in the $x$/$y$ direction. In this situation, we can see many phonon modes with large value of IPR, indicating localization. The largest value for IPR is about 0.2, which means that this mode is localized around about five atoms. In panel (b), we only consider the FNN interaction. There will be no coupling between $x$ and $y$ directions in the square lattice system. So condition (I) will be broken, yet the condition (II) still holds. In this situation, the $P^{-1}$ for all phonon modes are very small, on the order of $1/64=0.0156$. So there is no LEM. Panel (c) shows that the situation is similar if condition (II) is broken. In Fig~\ref{fig_rectangle}~(a)-(f), we show explicitly the six phonon modes denoted by labels a - f in Fig~\ref{fig_rectangle_ipr}. In panels (a) and (b), the first boundary atom has very large vibrational amplitude, and this vibrational amplitude decays exponentially from the edge into the center. The localization length is $L_{c}=0.9a$, where $a$ is the bond length. This very small localization length explicates the good localization property of the LEM. Actually, it can be seen from the figure that the vibrational amplitude is almost zero for the fourth atom from the edge. Panels (c)-(f) are the four non-localized phonon modes, where the first boundary atom can have large vibrational amplitudes, but the amplitudes do not decay.

In the above 2D square lattice system, the SNN interaction is required to introduce a coupling between
\begin{figure}[htpb]
  \begin{center}
    \scalebox{1.3}[1.2]{\includegraphics[width=7cm]{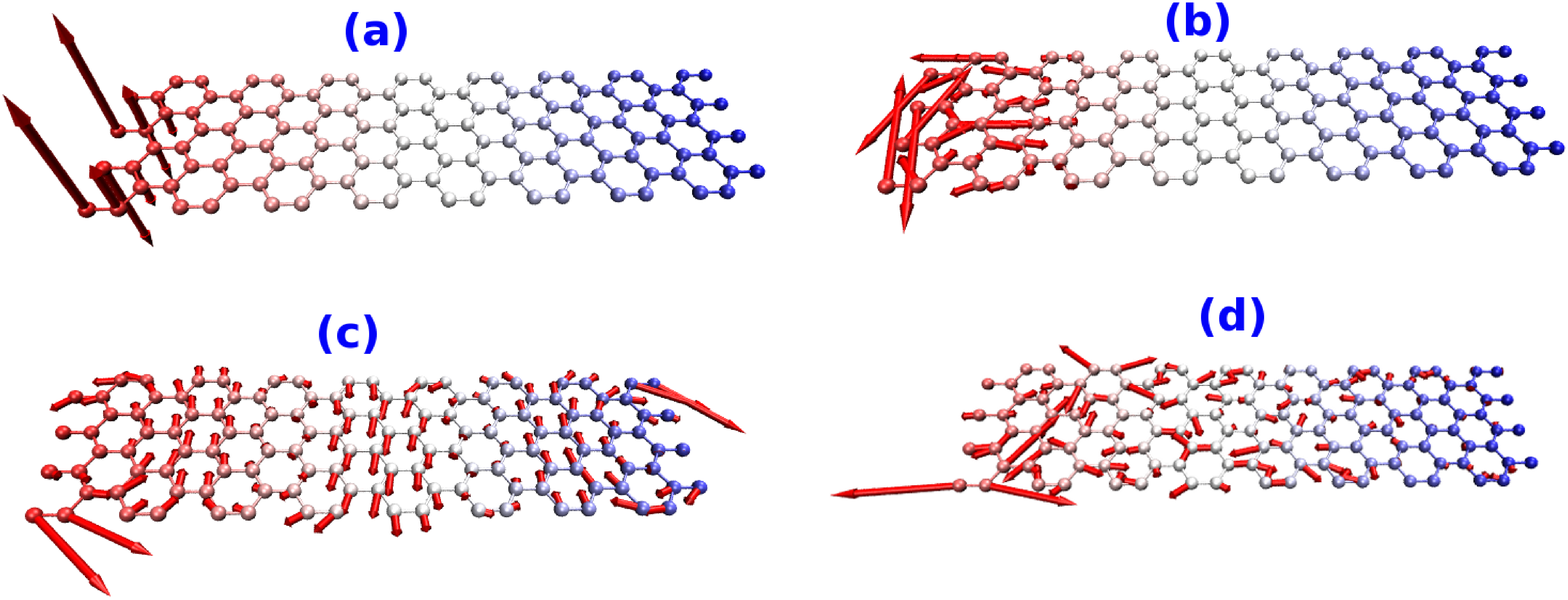}}
  \end{center}
  \caption{(Color online) LEM and non localized modes in graphene with Brenner potential. (a), (b) are two LEM and (c), (d) are two non localized modes.}
  \label{fig_graphene_Brenner}
\end{figure}
\begin{figure}[htpb]
  \begin{center}
    \scalebox{1.2}[1.1]{\includegraphics[width=7cm]{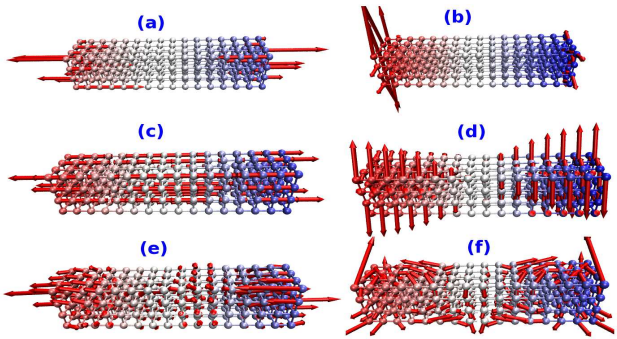}}
  \end{center}
  \caption{(Color online) The LEM in the simple cubic lattice system are shown in (a), (b). The non localized modes due the breaking of the condition (I), or (II) are in (c), (d) or (e), (f), respectively.}
  \label{fig_cubic}
\end{figure}
 $x$ and $y$ directions. Now we consider the 2D graphene sheet as shown in Fig.~\ref{fig_graphene}. The graphene lattice structure is quite different from the square lattice, as the FNN interaction is enough to couple $x$ and $y$ directions in the graphene. We first use the VFFM with the FNN interaction to describe the potential of the system. The open/periodic BC are applied in the $x$/$y$ directions, respectively. In this way, both conditions (I) and (II) are satisfied, and we indeed see the LEM as shown in panels (a) and (b). We can not break the condition (I) in graphene, since the satisfaction of condition (I) origins from the honeycomb structure. However, we can break the condition (II) by applying open BC in both $x$ and $y$ directions. Now there is no LEM in this system, and we can only see some extended modes represented by panels (c) and (d). We also apply the much more complicated Brenner potential to check whether the two conditions for the LEM is potential dependent or not. The carbon-carbon bond length is $a=1.42$~{\AA}. Fig.~\ref{fig_graphene_Brenner} gives us the same information as the VFFM potential in Fig.~\ref{fig_graphene}: if both conditions (I) and (II) are fulfilled, there is LEM as displayed in panels (a) and (b); otherwise, if the condition (II) is broken, there is only non-localized modes as shown in (c) and (d). This test tells us that the two conditions (I) and (II) for the LEM are very general. They do not dependent on the interaction potential of the system. Actually, we have also tested more potentials, such as Tersoff\cite{Tersoff} potential. We get the same result as Brenner potential. We have also considered the 2D triangular lattice structure. Similar to graphene, the condition (I) can be fulfilled by considering only the FNN interaction in the triangular lattice. We get the same result as graphene discussed here. We also note that the fixed BC prefers to have LEM compared with open BC. The LEM will always occur when fixed BC is applied in graphene.

In the above we have considered the low dimensional system. Now we are going to discuss the 3D system. We begin with the simplest 3D system: simple cubic lattice structure as displayed in Fig.~\ref{fig_cubic}. We use the VFFM interaction,
\begin{figure}[htpb]
  \begin{center}
    \scalebox{1.2}[1.1]{\includegraphics[width=7cm]{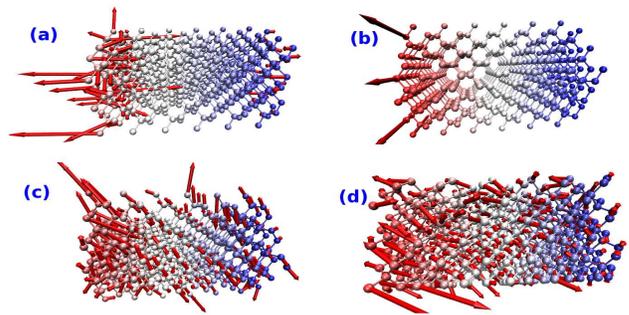}}
  \end{center}
  \caption{(Color online) The LEM and non localized modes in the diamond structure with VFFM vibrational potential. (a), (b) are two LEM. (c), (d) are two non localized modes obtained by breaking the condition (II).}
  \label{fig_diamond}
\end{figure}
 and consider up to the SNN interaction so that there is coupling between $x$, $y$ and $z$ directions, thus satisfying the condition (I). The condition (II) can also be satisfied by applying open BC in $x$ direction and periodic BC in $y$, $z$ directions. In this situation, we can see the LEM as shown in panels (a) and (b). We note that in this 3D system, the condition (II) does not require the application of periodic BC in both $y$ and $z$ directions. We only have to ensure that the periodic BC is applied to at least one of these two directions. If we only apply periodic BC in one direction (eg. in $y$ direction), we can also obtain the LEM, although the number of the LEM is smaller. If we break the condition (I) by considering only the FNN interaction, there are only the non-localized phonon modes as displayed in panels (c) and (d). If we break the condition (II) by applying open BC in all three directions, we can only get the non-localized phonon modes as demonstrated in panels (e) and (f).

We consider a more complicated 3D system, diamond lattice. The significant difference between the diamond structure and the simple cubic lattice structure is that the FNN interaction in diamond is enough to account for coupling between three directions. We use the VFFM with the FNN interaction to satisfy the condition (I);
\begin{figure}[htpb]
  \begin{center}
    \scalebox{1.1}[1.0]{\includegraphics[width=7cm]{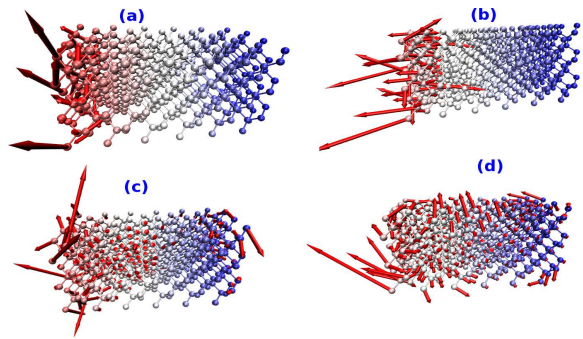}}
  \end{center}
  \caption{(Color online) The LEM and non localized modes in the diamond structure with Brenner potential. (a), (b) are two LEM. (c), (d) are two non localized modes due to the breaking of the condition (II).}
  \label{fig_diamond_Brenner}
\end{figure}
 and apply the open BC in the $x$ direction and periodic BC in both $y$ and $z$ directions to fulfill the condition (II). Under these two conditions, we can see the LEM as shown in Fig.~\ref{fig_diamond} (a) and (b). We can not break the condition (I) in the diamond structure, since this condition is naturally satisfied by its structure. We can break the condition (II) by applying open BC in all three directions. This leads to the non localized phonon modes as represented by (c) and (d). Besides the VFFM, we also use the Brenner potential to describe the interaction of the diamond with the bond length as $a=1.54$~{\AA}. Similar phenomena are shown in Fig.~\ref{fig_diamond_Brenner} where panels (a) and (b) are the LEM with both conditions (I) and (II) fulfilled. While panels (c) and (d) show two typical non localized phonon modes with condition (II) broken.

To conclude, we have used various examples to illustrate the two proposed conditions for the existence of the LEM: (I) coupling between different directions ($x$, $y$ or $z$) in the interaction; (II) different BC in three directions. The LEM will exist if these two conditions are satisfied simultaneously. We use different systems to show the generality of these two conditions: 1D chain, 2D square lattice, 2D graphene, 3D simple cubic, and 3D diamond lattice. These two conditions are also valid in case of different interaction potentials.

The present work sheds some light on how to efficiently make use of the LEM where it is beneficial, and avoid it when it causes a disaster. For example, in the study of thermal transport in real materials,\cite{Jiang} the potential has been fixed, so condition (I) can not be broken. We can apply same open BC in all directions as displayed in Fig.~\ref{fig_graphene} and Fig.~\ref{fig_graphene_Brenner}, to eliminate the LEM by breaking condition (II).

\textbf{Acknowledgements} The work is supported by a Faculty Research Grant of R-144-000-257-112 of National University of Singapore. We thank Dr. H. Tang for critically reading the manuscript.

\end{document}